\documentstyle[sprocl,epsfig]{article}

\def\Journal#1#2#3#4{{#1} {\bf #2}, #3 (#4)}

\def\APJ{\em Astrophys. J.}
\def\EPL{\em Europhys. Lett.}
\def\NCA{\em Nuovo Cimento}

\def\NPB{{\em Nucl. Phys.} B}
\def\NPBPS{{\em Nucl. Phys.} B (Proc. Suppl.)}
\def\PLB{{\em Phys. Lett.}  B}
\def\PPNP{\em Prog. in Part. and Nucl. Phys.}
\def\PREP{\em Phys. Rep.}
\def\PRL{\em Phys. Rev. Lett.}
\def\PRC{{\em Phys. Rev.} C}
\def\PRD{{\em Phys. Rev.} D}
\def\RMP{\em Rev. Mod. Phys.}
\def\SJNP{\em Sov. J. Nucl. Phys.}
\def\ZPC{{\em Z. Phys.} C}

\def\be{\begin{equation}}
\def\ee{\end{equation}}
\def\bea{\begin{eqnarray}}
\def\eea{\end{eqnarray}}

\def\gtwid{\mathrel{\raise.3ex\hbox{$>$\kern-.75em\lower1ex\hbox{$\sim$}}}}
\def\ltwid{\mathrel{\raise.3ex\hbox{$<$\kern-.75em\lower1ex\hbox{$\sim$}}}}

\begin{document}

\title{THE STATUS OF NEUTRINO MASS}

\author{DAVID O. CALDWELL}

\address{Institute for Nuclear and Particle\\
         Astrophysics and Cosmology and\\
         Physics Department, University of California,\\
         Santa Barbara, CA 93106-9530, USA}


\maketitle\abstracts{
New experimental results, if correct, require at least one light sterile
neutrino, in addition to the three active ones, to accommodate the mass
differences required to explain the solar $\nu_e$ deficit, the anomalous
$\nu/e$ ratio produced by atmospheric neutrinos, and either the candidate
events for $\nu_\mu\to\nu_e$ (or $\bar\nu_\mu\to\bar\nu_e$) from the LSND
experiment, or the possible need for a hot component of dark matter.  This
neutrino mass pattern can not only accommodate all these four requirements, but
also provide a robust solution to a problem presently making heavy-element
synthesis by supernovae impossible and resolve a possible discrepancy 
between big bang nucleosynthesis theory and observations.}
  
\section{Introduction}

The evidence is now becoming very strong for a particular pattern of neutrino
masses, one which requires at least one light sterile neutrino.  Either some of
the experimental results are wrong, or we are forced to this conclusion.  This
experimental evidence will be reviewed briefly, with emphasis on very new
results, and the consequences for neutrino mass examined.  Indirect evidence
for the same mass pattern from dark matter, supernova nucleosynthesis, and
big-bang nucleosynthesis will then be presented, showing the widespread effect
of massive neutrinos.  This mass pattern should, however, produce some
apparently negative results from current experiments which could have a
deleterious effect on the field unless they are properly anticipated.

\section{Indications for Nonzero Neutrino Mass}
\subsection{Solar Neutrino Deficit}

\begin{figure}
\epsfxsize=12cm
\epsfbox{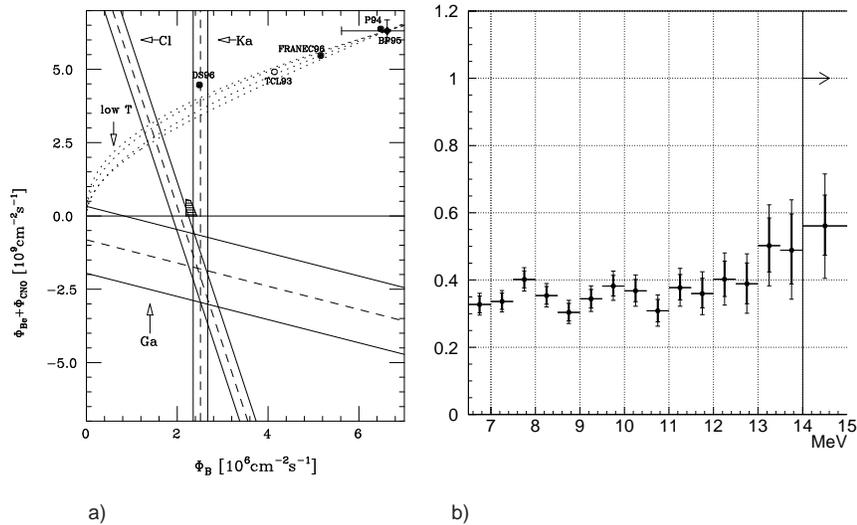}
\caption{a). The $^8$B and $^7$Be (+ CNO) neutrino fluxes for standard neutrinos. 
The dashed (solid) lines correspond to central ($\pm1\sigma$) experimental
values for the chlorine (Cl), gallium (Ga) and Kamiokande (Ka) experiments. 
The hatched area corresponds to a region within $2\sigma$ from each
experimental result.  The predictions of solar
models are shown, with the one\protect\cite{ref:7} 
with error bars being most often referenced. The dotted lines indicate the
behavior of non-standard solar models with low central temperature.
b).The Super-Kamiokande solar neutrino energy spectrum with dark error
bars for statistics and light error bars including systematics, mainly
determined by the energy calibration.
\label{fig:1}}
\end{figure}

All solar neutrino experiments observe fewer electron neutrinos ($\nu_e$) than
solar models predict.  In addition, because the three types of experiments
cover different $\nu_e$ energy ranges and hence sample differently the
contributions from the various nuclear processes producing neutrinos, there is
an energy-dependent discrepancy well illustrated in Fig.~\ref{fig:1}(a).  This
figure, from a very complete review,\cite{ref:1} shows the relationship between
neutrino fluxes from $^7$Be and $^8$B neutrinos as measured in the three types
of experiments.  The SAGE\hphantom{,}\cite{ref:2} and
GALLEX\hphantom{,}\cite{ref:3} radiochemical experiments go to the lowest
energy and hence measure all of both fluxes (designated ``Ga"), while the
Homestake\hphantom{,}\cite{ref:4} radiochemical experiment measures all of the
$^8$B spectrum but only part of the $^7$Be flux (labeled as ``Cl"), and the
Kamiokande\hphantom{,}\cite{ref:5} and Super-Kamiokande\hphantom{,}\cite{ref:6}
scattering experiments measure only $^8$B flux (designated as ``Kam").  Results
from all three actually intersect at a negative value of the $^7$Be flux, yet
$^8$B is produced from $\rm^7Be+p\to\/^8B+\gamma$.  This problem cannot be
avoided by one of the experiments being wrong.  The discrepancy between a
standard solar model\hphantom{,}\cite{ref:7} and all three types of experiments
is shown by the point with error bars in the upper right-hand corner
indicating predicted fluxes.  Solar models which drastically change solar
properties do not solve the problem.  Recent very
accurate helioseismology measurements severely constrain solar models and
apparently rule out any astrophysical explanation\hphantom{,}\cite{ref:8} of
the solar neutrino discrepancies.

A good solution to the solar $\nu_e$ deficit is provided by oscillation into
$\nu_\mu$, $\nu_\tau$, or $\nu_s$, a sterile neutrino, one not having the
normal weak interactions.  While this can be a vacuum oscillation, requiring a
mass-squared difference $\Delta m^2\sim10^{-10}$ eV$^2$ and large mixing
between $\nu_e$ and the other neutrino, more favored is a matter-enhanced
MSW\hphantom{,}\cite{ref:9} type of oscillation.  For a $\nu_\mu$ or $\nu_\tau$
final state, $\Delta m^2_{ei}\sim10^{-5}$ eV$^2$ and mixings either
$\sin^22\theta_{ei}\sim6\times10^{-3}$ or $\sim0.6$ are possible, while only
the former is allowed for $\nu_s$. The main change as a result of the new 
Super-Kamiokande data is that the
lack of a day-night effect has reduced the parameter space for the large-angle
solution for the $\nu_\mu$ or $\nu_\tau$ final state.~\cite{ref:10}  The Super-Kamiokande
result which will become of prime importance as the error bars are reduced is
shown in Fig.~\ref{fig:1}(b).  This energy spectrum could not only choose among
the oscillation solutions---and it is well fit at this stage by the MSW
small-angle solution---but also it may be the one means of proving that
oscillations are occurring, as will be explained later.

\subsection{Atmospheric Neutrino Deficit}

\begin{figure}
\epsfxsize=12cm
\epsfbox{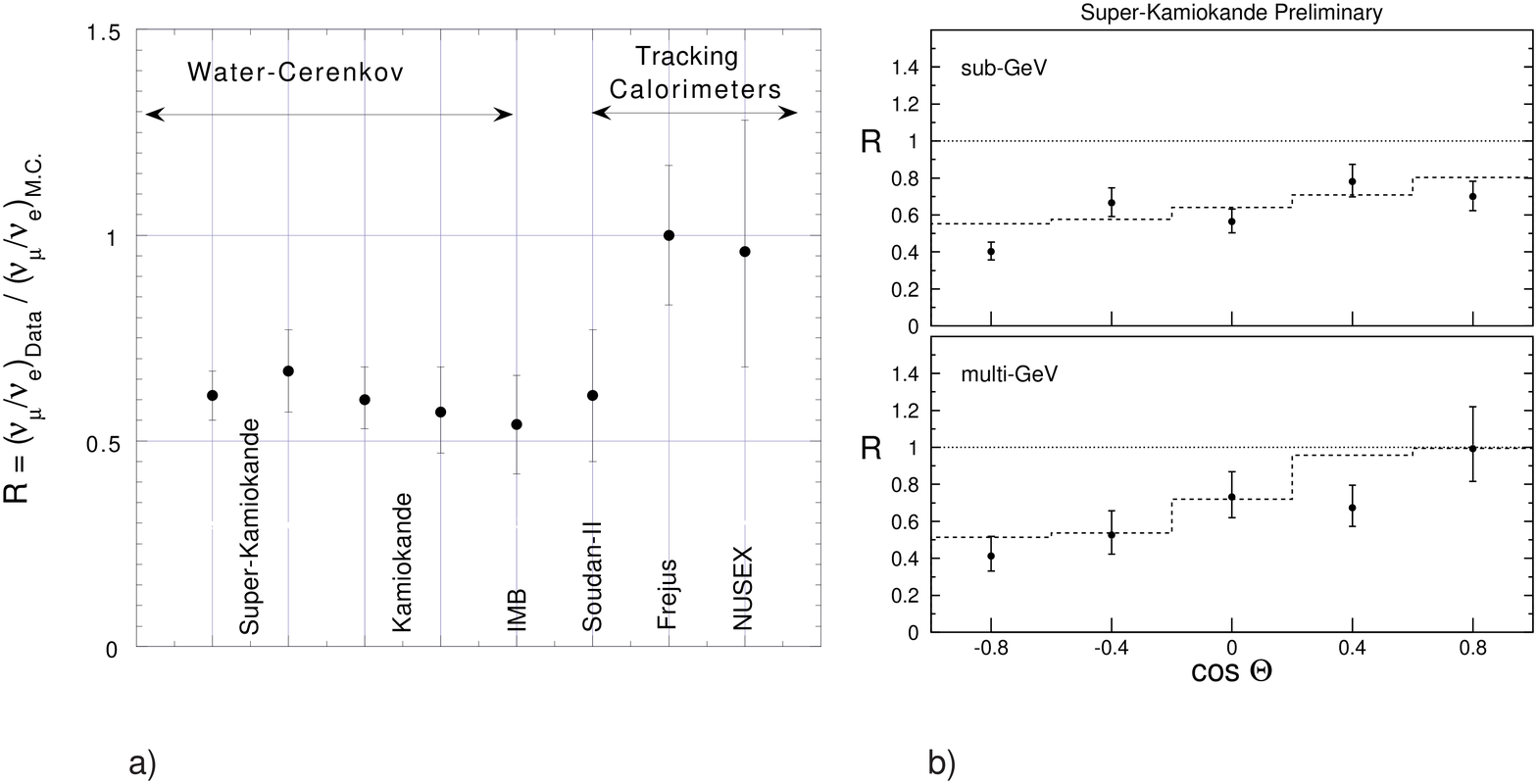}
\caption{a) The double ratio of $\nu_\mu/\nu_e$ from data to that from
calculations for atmospheric neutrinos for various detectors.  Note that
Super-Kamiokande and Kamiokande each have two independent data sets, above and
below 1.3 GeV; b) The ratio $(\mu/e)_{\rm DATA}/(\mu/e)_{\rm MC}$ for sub-GeV and
multi-GeV atmospheric neutrino samples from Super-Kamiokande, as a function of
zenith angle.  Neutrinos coming from below are at $\cos\Theta=-1$.  The dashed
line shows the expected shape for $\nu_\mu\to\nu_\tau$ oscillation with
$\sin^22\theta=1$ and $\Delta m^2=.005$ eV$^2$.
\label{fig:4}}
\end{figure}

Pions produced in the atmosphere would decay via $\pi\to\mu+\nu_\mu,\/\mu\to
e+\nu_\mu+\nu_e$, so that one would expect
$N(\nu_\mu+\bar\nu_\mu)=2N(\nu_e+\bar\nu_e)$, with a small correction for $K$
decays.  The $(\nu_\mu+\bar\nu_\mu)/(\nu_e+\bar\nu_e)$ ratio would be observed
in underground experiments as $\mu^\pm/e^\pm$, and the result is far from the
expected value.  Because the calculated $\mu^\pm$ and $e^\pm$ individual fluxes
are known to $\sim15$\%, whereas much of the uncertainty drops out in the
ratio, the experiments utilize $R=(\mu/e)_{\rm Data}/(\mu/e)_{\rm Calc}$. 
Values of $R$ for many experiments are shown in Fig.~\ref{fig:4}(a).  While it
once appeared that there was a discrepancy between water Cherenkov detectors
and tracking calorimeters,\cite{ref:11} the Soudan II
results\hphantom{,}\cite{ref:12} agree with those from IMB,\cite{ref:13}
Kamiokande,\cite{ref:14} and Super-Kamiokande.\cite{ref:15}  In addition, the
MACRO detector finds a similar deficiency of muons, although their angular
distribution is ambiguous, slightly favoring a neutrino oscillation explanation
of the lack of muons.\cite{ref:16}

While the statistical evidence for $R$ being less than unity is now quite
compelling, it is the angular distributions of
the $\mu$ and $e$ events which provide the primary evidence that this deviation
of $R$ from unity is explained by neutrino oscillations.  This non-flat
distribution with angle of $R$ was first observed in the high-energy ($>1.3$
GeV) event sample from Kamiokande, but has now been confirmed with better
statistics in the similar data sample from Super-Kamiokande, as shown in
Fig.~\ref{fig:4}(b).  The data fits an oscillation hypothesis, using $\Delta
m^2=5\times10^{-3}$ eV$^2$, $\sin^22\theta=1$ (as a sample, but not a best,
fit) and is far from a non-oscillation, flat distribution.  The low-energy
($<1.3$ GeV) sample also agrees with the same oscillation parameters, but this
should be a much shallower angle dependence, and hence it is statistically less
compelling, as is also shown in the figure.

\begin{figure}
\epsfxsize=12cm
\epsfbox{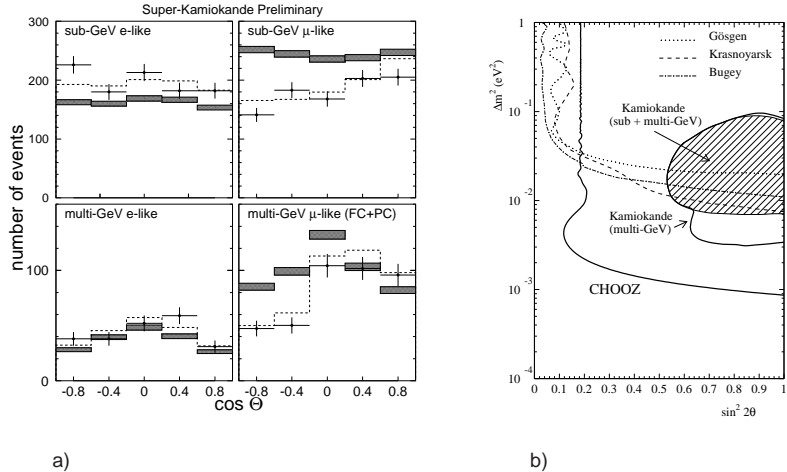}
\caption{a) The rate of $\mu$-like and $e$-like events in the sub-GeV and
multi-GeV atmospheric neutrino samples from Super-Kamiokande, as a function of
zenith angle.  The solid histograms are the Monte Carlo expectation with no
neutrino oscillation; the thickness represents the statistical uncertainty in
the Monte Carlo sample.  The dashed line shows the expected shape for
$\nu_\mu\to\nu_\tau$ oscillation with $\sin^22\theta=1$ and $\Delta
m^2=.005$ eV$^2$. b)The 90\% C.L.\ exclusion plot for CHOOZ, compared with previous
experimental limits and with the Kamiokande allowed region for $\nu_e$
disappearance.
\label{fig:6}}
\end{figure}

The disappearance of the muon neutrinos could be due to $\nu_\mu\to\nu_\tau$ or
$\nu_\mu\to\nu_e$, with $\nu_\mu\to\nu_s$ being unlikely because the large
mixing angle would bring the $\nu_s$ into equilibrium in the early universe,
possibly providing too many neutrinos to get agreement between predictions of
nucleosynthesis and observed light element abundances.  The Super-Kamiokande
observations of $e$ and $\mu$ compared to calculated fluxes, as well as the
individual $e$ and $\mu$ angular distributions, as shown in Fig.~\ref{fig:6}(a),
makes $\nu_\mu\to\nu_e$ very unlikely.  Note the $e$ distributions are like the
non-oscillation Monte Carlo, whereas those for $\mu$ agree with the oscillation
prediction.  The recent results of the CHOOZ nuclear reactor
experiment,\cite{ref:17} shown in Fig.~\ref{fig:6}(b), which does not see
evidence of $\nu_e$ disappearing in the appropriate region of $\Delta m^2$ and
$\sin^22\theta$, confirms that the atmospheric effect is very unlikely to be 
$\nu_\mu\to\nu_e$.
On the basis that the Super-Kamiokande observed values of $R$ and angular
distributions of $R$ are due to $\nu_\mu\to\nu_\tau$, the likely value of 
$\Delta m^2$ is definitely much larger than that
required for an explanation of the solar neutrino deficit, and the flavors
of neutrinos cannot be the same in the two cases.  Turning now to the third
possible manifestation of neutrino mass, we shall see that the atmospheric
$\Delta m^2$ is much smaller than that required for the LSND experiment, and
hence that three distinctly different values of neutrino mass differences are
required.

\subsection{Evidence for Neutrino Oscillations from the LSND Experiment}

The LSND accelerator experiment uses a decay-in-flight $\nu_\mu$ beam of up to
$\sim180$ MeV from $\pi^+\to\mu^+\nu_\mu$ and a decay-at-rest $\bar\nu_\mu$
beam of less than 53 MeV from the subsequent $\mu^+\to e^+\nu_e\bar\nu_\mu$. 
The 1993+1994+1995 data sets included 22 events of the type $\bar\nu_ep\to
e^+n$, based on identifying an electron between 36 and 60 MeV using Cherenkov
and scintillation light and tightly correlated with a $\gamma$ ($<0.6$\%
accidental rate) from $np\to d\gamma$ (2.2 MeV), whereas only $4.6\pm0.6$ such
events were expected from backgrounds.\cite{ref:18}  The chance that these
data, using a water target, result from a fluctuation is $<10^{-7}$. 
Subsequent data sets from 1996+1997 taken with an iron target gave a similar
oscillation probability with much worse statistical accuracy.  More
importantly, the first data sets (1993--5) yielded events from 
$\pi$ decay in flight consistent with being
from $\nu_\mu\to\nu_e$. These were similar in number to those from
$\bar\nu_\mu\to\bar\nu_e$, but with about twice the background, since the
observed process ($\nu_eC\to e^-X$) gave only one signal instead of two.  While
the fluctuation probability in this case is only $\sim10^{-2}$, the two ways of
detecting oscillations are essentially independent.\cite{ref:19}

\begin{figure}
\epsfxsize=12cm
\epsfbox{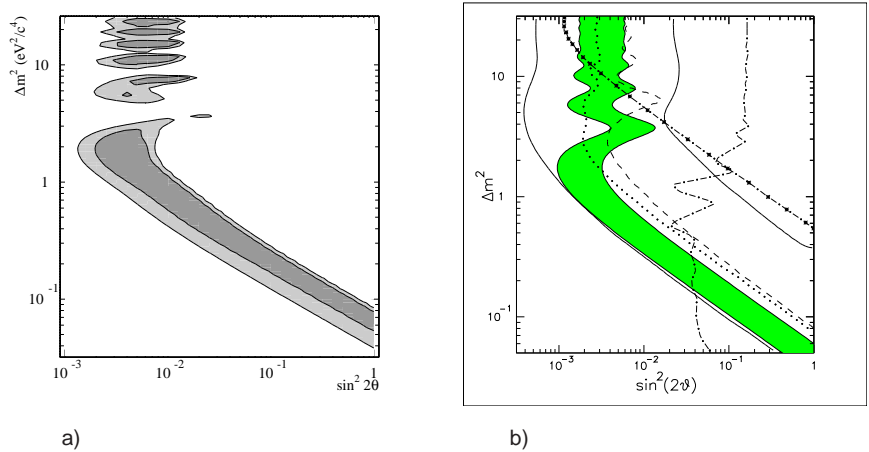}
\caption{a) Mass-squared difference ($\Delta m^2$) vs.\ degree of mixing
($\sin^22\theta$) assuming a two-neutrino oscillation explanation of the LSND
beam-excess data.  Shown are regions of $\Delta m^2$ favored using the energy
(from 20 to 60 MeV) and distance from the source of each event.b) As in (a) 
but the LSND $\bar\nu_\mu$ data here
provide an 80\% C.L.\ band, which is to be compared with the LSND $\nu_\mu$
result (solid lines), KARMEN (dashes), E776 (dots), Bugey (dash-dot), and NOMAD.
\label{fig:9}}
\end{figure}

While the $\nu_\mu\to\nu_e$ results are consistent with those from
$\bar\nu_\mu\to\bar\nu_e$, only the latter have sufficient statistics to
provide restrictions on the value of $\Delta m^2$.  These $\bar\nu_\mu$ results
interpreted as a two-generation oscillation have been
presented\hphantom{,}\cite{ref:18} in a plot like Fig.~\ref{fig:9}(a), except that
comparisons were made to limits from other experiments.  Figure~\ref{fig:9}(a) is
the correct way to determine favored regions of $\Delta m^2$ as a function of
the mixing angle, $\theta$.  The plot utilizes all the information about the
events, in particular the neutrino energy, $E$, and the distance of the event
from the source, $L$.  In order to increase the range of $L/E$, values of $E$
down to 20~MeV were used.  Figure~\ref{fig:9}(a) shows contours at 2.3 and 4.5
log-likelihood units from the maximum.  If this were a gaussian distribution,
which it is not (its integral being infinite), the contours would correspond to
90\% and 99\% likelihood levels, but in addition they have been smeared to
account for some systematic errors.  Comparison to the KARMEN
experiment,\cite{ref:20} which presents results in a similar way, shows no
conflict, but if limits are plotted (as they are in Ref.~18) on this graph from
E776 at BNL\hphantom{,}\cite{ref:21} and the Bugey reactor
experiment,\cite{ref:22} then one might conclude that the only allowed $\Delta
m^2$ region is 0.2--3 eV$^2$.  If instead an 80\% confidence level band is
calculated to compare with the 90\% confidence level limits of those
experiments using, as they do, just numbers of events (i.e., not using the L/E
information) and using only the 36--60 MeV region with its much lower
background, then there is no conflict with other experiments above 0.2 eV$^2$,
up to the recent limit of about 10 eV$^2$ from the NOMAD
experiment,\cite{ref:23} as shown in Fig.~\ref{fig:9}(b).

\section{Pattern of Neutrino Masses Required by Experiments}

Because measurements of the width of the $Z^0$ boson require that there be
only three light neutrinos coupled to the $Z^0$, it would be desirable to
explain the phenomena described in the previous section in terms of
oscillations among those three neutrinos.  Since the flavors are constrained,
one has to invoke indirect neutrino oscillations, so LSND could be observing
$\nu_\mu\to\nu_\tau\to\nu_e$, for example, with the largest (dominant) $\Delta
m^2$ being between $\nu_\mu$ and $\nu_\tau$ or $\nu_e$ and $\nu_\tau$, with a
small $\Delta m^2_{e\mu}$.  There is still the problem that three neutrinos
provide only two mass-squared differences, so one might assume $\Delta m^2_{\rm
solar}\approx\Delta m^2_{\rm atmos.}$, as did Acker and Pakvasa,\cite{ref:24}
requiring both processes to be dominantly $\nu_e\rightleftharpoons\nu_\mu$. 
This leads to requiring the solar $\nu_e$ deficit to be energy independent, in
conflict with the data (see, e.g., Fig~\ref{fig:1}).  This is
a problem besetting most three-neutrino schemes.  The Acker-Pakvasa model is
essentially ruled out by the CHOOZ result of Fig.~\ref{fig:6}(b), as well as the
angular distributions of Fig.~\ref{fig:6}(a).

The only other alternative for three neutrinos is making $\Delta m^2_{\rm
atmos.}=\Delta m^2_{\rm LSND}$.  This was suggested by Cardall and
Fuller,\cite{ref:25} who used the indirect oscillation for LSND and made
$\Delta m^2_{e\tau}\approx\Delta m^2_{\mu\tau}\approx0.3$ eV$^2$, since the
solar $\Delta m^2_{e\mu}\approx10^{-5}$ eV$^2$ is so small.  This scheme has
no difficulty with the solar data, but could be in some conflict with limits
from neutrinoless double beta decay, if one wants to provide a hot dark matter
component, since $m_{\nu_e}\approx m_{\nu_\mu}\approx m_{\nu_\tau}\approx1.6$
eV, as will be explained in the next section.  This scheme, the three-neutrino
pattern having the least conflict with data,\cite{ref:26} is definitely ruled
out by the Super-Kamiokande data shown in Figs.~\ref{fig:4}(b) and 
\ref{fig:6}(a).  There is no way the $\Delta m^2$ required by LSND can be the same
as that needed for the atmospheric anomaly.

Either the experiments are wrong or one is forced to invoke another light
neutrino, a sterile one which does not have the normal weak interactions and
hence does not couple to the $Z^0$.  Then the solar $\nu_e$ deficit is a result
of $\nu_e\to\nu_s$, with $\Delta m^2_{es}\ltwid10^{-5}$ eV$^2$, the
atmospheric $\nu_\mu/\nu_e$ ratio is explained by $\nu_\mu\to\nu_\tau$, with
$\Delta m^2_{\mu\tau}\sim10^{-2}-10^{-3}$ eV$^2$, and the LSND observation is
caused by $\nu_\mu\to\nu_e$, with $0.2\ {\rm eV}^2\ltwid\Delta
m^2_{e\mu}\ltwid10$ eV$^2$.  The information considered so far requires this
and only this, but we shall see that a wide range of phenomena can be explained
for a more specified value of $\Delta m^2_{e\mu}$ and for which the sterile
neutrino is a necessity.

\section{Implications of the Four-Neutrino Mass Pattern}

\subsection{\label{sec:4.1}Dark Matter}

This four-neutrino mass pattern was first proposed\hphantom{,}\cite{ref:27}
five years ago, well before there were any results from LSND.  The motivation
then was to explain the solar and atmospheric deficits and the apparent need
for some of the missing mass of the universe to be in the form of neutrinos. 
The neutrino mass required is $94\ h^2\Omega F_\nu$, where $h$ is the Hubble
constant in units of 100 km$\cdot$s$^{-1}\cdot$Mpc$^{-1}$, $\Omega$ is the
density of the universe in units of the critical density, and $F_\nu$ is the
neutrino fraction of $\Omega$.  For example, if $h=0.5$, $\Omega=1$, and
$F_\nu=0.2$, then 4.7 eV of neutrino mass is needed.  Before LSND it was
suggested\hphantom{,}\cite{ref:27} that there were only two ways to get the
needed neutrino mass as well as explain the solar and atmospheric deficits: 1)
the four-neutrino scheme with $\nu_e$ and $\nu_s$ light ($\ll1$ eV), and
$\nu_\mu$ and $\nu_\tau$ sharing the dark matter role ($\sim2$--3 eV each), or
2) $\nu_e\to\nu_\mu$ for solar, $\nu_\mu\to\nu_\tau$ for atmospheric, and all
three neutrinos providing dark matter $(\sim1.6$ eV each).  After LSND the
latter scheme failed the flavor constraint, so indirect oscillations had to be
invoked, and it became the Cardall-Fuller proposal.\cite{ref:25}  Now only the
former (four-neutrino) scheme remains.  The main point of mentioning this
history is to emphasize that the four-neutrino pattern can be motivated even
without LSND.

The dark matter part of that motivation arises because relic neutrinos remain
($\sim100/\rm cm^3$ per neutrino flavor) from the early universe, and if they
have even a few eV in mass they could solve the main problem of cold dark
matter (CDM) models, namely production of much more structure on small scales
than is observed.  This results from baryons being readily accreted to
overdense regions of the slowly moving CDM.  CDM models give a quite good
approximation to the structure of the universe over a wide range of distance
scales, but when an absolute normalization of the predictions was provided by
the COBE data, the overproduction of small-scale structure became apparent. 
Since the free streaming of neutrinos could reduce density fluctuations on
small scales, the addition of 30\% neutrinos (i.e., $F_\nu=0.3$) allowed
fitting structure on all scales very well.  The problem with this early
cold+hot dark matter model\hphantom{,}\cite{ref:28} was that this damping of
density perturbations also caused structure to form too late.  Reducing the
neutrino content to $\sim20$\% allowed early enough structure
formation.\cite{ref:29}  With all the mass (4.7 eV) in one neutrino species,
this otherwise successful model (C$\nu$DM) overproduced clusters of galaxies. 
In other words, the C$\nu$DM model worked well at all distance scales except
$\sim10h^{-1}$ Mpc.  With the motivation of the four-neutrino model discussed
above, simulations were tried in which the dark matter was shared between two
2.4 eV neutrinos, yielding a quite unexpected result.\cite{ref:30}  While 4.7
eV in one neutrino species or two makes essentially no difference at very large
or very small scales, at $\sim10h^{-1}$ Mpc the larger free-streaming length of
the 2.4 eV neutrinos washes out density fluctuations and hence lowers the
abundance of galactic clusters.

This C$\nu^2$DM model with two, 2.4 eV neutrinos fits structure information on
all scales.  In every aspect of simulations done subsequently the two-neutrino
dark matter gives the best results.  For example, a single neutrino species (as
well as low-$\Omega$ models) overproduce void regions between galaxies, whereas
the C$\nu^2$DM model agrees with observations.\cite{ref:31}  Note that the
C$\nu^2$DM model is compatible with all the information mentioned in Section 2
provided $\Delta m^2_{e\mu}$ from LSND is $\sim6$--8 eV$^2$.  This happens to
be the region of Fig.~\ref{fig:9}(a) corresponding to the second oscillation
maximum for $\bar\nu_\mu\to\bar\nu_e$ and the first oscillation maximum
(Fig.~\ref{fig:9}(b)) for $\nu_\mu\to\nu_e$, so if this is the correct $\Delta
m^2$ the target-to-detector distance for LSND is extremely fortuitous.

The C$\nu^2$DM model works if $\Omega=1$ and $h\ltwid0.6$.  Not long ago
large values of $h$ were popular, providing a universe age crisis.  Formerly
high values of $h$ have been reduced, and the Hipparcos satellite
measurements\hphantom{,}\cite{ref:32} of stellar parallaxes reduced both the
age of the universe and $h$ values.  Now there is no conflict between the age
of the oldest stars and the currently favored $h\sim0.6$, with many
measurements of $h$ coming out even lower than that.

The latest bandwagon is a low-$\Omega$ universe, contrary to the expectation of
almost all models of a period of inflationary expansion of the universe which
require $\Omega=1$ and explain the isotropy of the cosmic microwave background,
the flatness problem, lack of monopoles, and the origin of large-scale
structure.  To have a density not now near zero or infinity requires
$\Omega+\Lambda=1$, where $\Lambda$ is an arbitrary cosmological constant
having little or no theoretical justification.  One reason low-$\Omega$ is
popular is based on assuming that galactic clusters are representative of the
universe as a whole, using X-ray measurements of the gas in clusters to
determine the fraction of mass in baryons, $F_B$, and taking the value of the
baryon density in the universe, $\Omega_B$, as determined by primordial He
abundance to get $\Omega=\Omega_B/F_B=0.05/0.15=0.3$.  The X-ray measurements,
depending on a collision process, could give an overestimate if the gas is
clumped, so that recent measurements of the Sunyaev-Zeldovitch
effect\hphantom{,}\cite{ref:33} are more reliable and give
$F_B=(0.06\pm0.01)h^{-1}$.  A particularly sensitive determination of the
baryon-to-photon ratio at the time of nucleosynthesis is provided by the
primordial deuterium to hydrogen ratio, D/H.  There were conflicting values of
D/H in very high red-shift clouds, but these are now resolved in favor of low
D/H,\cite{ref:34a} from which one gets
$\Omega_B=(0.024^{+0.006}_{-0.005})h^{-2}$.  Then $\Omega=0.4h^{-1}$, so
$\Omega=0.7$--0.8, quite consistent with $\Omega=1$, since a large-scale
simulation\hphantom{,}\cite{ref:34} of measured galactic cluster properties
yields values like $\Omega=0.5$ when the input to the simulation is $\Omega=1$. 
This discrepancy could explain some other observations apparently favoring low
$\Omega$.

Particularly important to note is that if the LSND measurement is confirmed by
another experiment which finds $\Delta m^2_{e\mu}\sim6$--8 eV$^2$, the
existence of one 2--3 eV neutrino would require large $\Omega$.\cite{ref:30} 
Such a neutrino washing out density fluctuations in the early universe would
not allow sufficient structure to form for low $\Omega$, making $\Omega=1$
highly probable.  Thus a neutrino experiment could settle the much-disputed
issue of the ultimate fate of the universe.

\subsection{Heavy-Element Nucleosynthesis in Supernovae}

While the $\Delta m^2_{e\mu}\sim6$--8 eV$^2$ value is necessary for the
successful two-neutrino dark matter, it causes an apparent conflict with the
production of heavy elements in supernovae.  This $r$-process of rapid neutron
capture occurs in the outer neutrino-heated ejecta of Type II supernovae.  The
existence of this process would seem to place a limit on the mixing of
$\nu_\mu$ and $\nu_e$ because energetic $\nu_\mu\ (\langle E\rangle\approx25$
MeV) coming from deep in the supernova core could convert via an MSW transition
to $\nu_e$ inside the region of the $r$-process, producing $\nu_e$ of much
higher energy than the thermal $\nu_e\ (\langle E\rangle\approx11$ MeV).  The
latter, because of their charge-current interactions, emerge from farther out
in the supernova where it is cooler.  Since the cross section for $\nu_en\to
e^-p$ rises as the square of the energy, these converted energetic $\nu_e$
would deplete neutrons, stopping the $r$-process. 
Calculations\hphantom{,}\cite{ref:35} of this effect limit $\sin^22\theta$ for
$\nu_\mu\to\nu_e$ to $\ltwid10^{-4}$ for $\Delta m^2_{e\mu}\gtwid2$ eV$^2$, in
conflict with compatibility between the LSND result and a neutrino component of
dark matter.

The sterile neutrino, however, can not only solve this problem, but also rescue
the $r$-process itself.  While recent simulations have found the $r$-process
region to be insufficiently neutron rich, very recent realization of the full
effect of $\alpha$-particle formation has created a disaster for the
$r$-process.\cite{ref:36}  The initial difficulty of too low entropy (i.e., too
few neutrons per seed nucleus, like iron) has now been drastically exacerbated
by calculations\hphantom{,}\cite{ref:36} of the sequence in which all available
protons swallow up neutrons to form $\alpha$ particles, following which
$\nu_en\to e^-p$ reactions create more protons, creating more $\alpha$
particles, and so on.  The depletion of neutrons by making $\alpha$ particles
and by $\nu_en\to e^-p$ rapidly shuts off the $r$-process, and essentially no
nuclei above $A=95$ are produced.

The sterile neutrino would produce two effects.\cite{ref:37}  First, there is
a zone, outside the neutrinosphere (where neutrinos can readily escape) but
inside the $\nu_\mu\to\nu_e$ MSW (``LSND") region, where the $\nu_\mu$
interaction potential goes to zero, so a $\nu_\mu\to\nu_s$ transition can occur
nearby, depleting the dangerous high-energy $\nu_\mu$ population.  Second,
because of this $\nu_\mu$ reduction, the dominant process in the MSW region
reverses, becoming $\nu_e\to\nu_\mu$, dropping the $\nu_e$ flux going into the
$r$-process region, hence reducing $\nu_en\to e^-p$ reactions and allowing the
region to be sufficiently neutron rich.  This
rescuing scenario---the only robust one which has been found after many
attempts---works even better if the MSW region is inside the radius at which the
weak interactions freeze out.  This density requirement is well satisfied for
$\Delta m^2_{e\mu}\sim6$ eV$^2$, a value which cannot be reduced appreciably.

\subsection{Light-Element Nucleosynthesis in the Early Universe}

Another possible indication for the existence of the sterile neutrino comes
from the big-bang production of light elements.  The apparent incompatibility
of determinations of the baryon-to-photon ratio (hence fraction of baryons or
the number of light degrees of freedom, like neutrinos) on the basis of $^4$He
abundance as opposed to the low D/H ratio, indirectly alluded to in
Sec.~\ref{sec:4.1}, could also be resolved by the sterile neutrino.  While not
the most precise way of doing so, this issue is more easily discussed in terms
of the effective number of light neutrinos, $N_{\rm eff}$, in equilibrium at
the time of nucleosynthesis.  To reconcile the deduced primordial $^4$He
abundance with the low D/H values a universe expansion rate at the time of
decoupling of the neutron-to-proton ratio would have to be governed by $N_{\rm
eff}=1.9\pm0.3$, or put another way, standard big-bang nucleosynthesis is said
to be excluded at the 99.9\% C.L.\cite{ref:38}  Adding sterile neutrinos would
seem only to make the problem worse.  Foot and Volkas have
suggested,\cite{ref:39} however, that the lepton number asymmetry created by
transitions of $\bar\nu\to\bar\nu_s$ could lead to a significant excess of
$\nu_e$ over $\bar\nu_e$, so that the $n/p$ ratio would be depleted prior to
the decoupling of the $\nu_en\to e^-p$ reaction, leading to the production of
less $^4$He.  This would produce the same result as artificially changing the
universe expansion rate by making $N_{\rm eff}<3$.  This lepton asymmetry comes
about because the conditions for a given MSW transition will not produce both
$\nu\to\nu_s$ and $\bar\nu\to\bar\nu_s$, and the scarceness of initial sterile
neutrinos makes the dominant MSW transition active$\to$sterile and not the
other way around.

The small $\Delta m^2$ of the solar case makes $\bar\nu_e\to\bar\nu_s$
have a negligible effect.  However, $\bar\nu_\mu\to\bar\nu_s$ and/or
$\bar\nu_\tau\to\bar\nu_s$ with $\Delta m^2\sim6$--8 eV$^2$ could create a
large lepton asymmetry which can be transformed to $\nu_e$ through
$\nu_\mu\to\nu_e$ and/or $\nu_\tau\to\nu_e$.  Foot and Volkas
find\hphantom{,}\cite{ref:39} that $\Delta m^2>3$ eV$^2$ is required since at a
lesser $\Delta m^2$ the MSW density would be achieved at too late a time.  This
restriction is very like that of the supernova nucleosynthesis case, again
hinting that the LSND result lies in the upper range of the allowed region. 
There is also the interesting possibility that nearly
mass-degenerate $\nu_\mu$ and $\nu_\tau$ could produce about double the effect Foot
and Volkas find, $\delta N_{\rm eff}=-0.5$, and hence 3 active neutrinos could
give $N_{\rm eff}\approx 2$ for the 4-neutrino scheme described above.

\section{Conclusions}

Either one of the three experimental evidences for neutrino oscillations (solar
$\nu_e$ deficit, anomalous atmospheric $\nu_\mu/\nu_e$ ratio, and LSND events,
or as an alternative to the last, the need for a neutrino component of dark
matter) is wrong or a neutrino mass pattern is required which includes at least
one light sterile neutrino.  This pattern utilizes $\nu_e\to\nu_s$ for the
solar effect with $\Delta m^2_{es}\ltwid10^{-5}$ eV$^2$, $\nu_\mu\to\nu_\tau$
for the atmospheric case with $\Delta m^2_{\mu\tau}\sim10^{-2}$--$10^{-3}$
eV$^2$, and $\nu_\mu\to\nu_e$ for LSND's events with $0.2<\Delta m^2_{e\mu}<10$
eV$^2$.  If in addition the $\nu_e$ and $\nu_s$ are $\ll1$ eV and the $\nu_\mu$
and $\nu_\tau$ are $\sim2.4$ eV each (so that $\Delta m^2_{e\mu}\sim6$--8
eV$^2$) then this pattern also provides the best hot+cold dark matter model,
and it fits universe structure on all scales.  The $\nu_s$ provides the only
known robust solution to an otherwise disasterous failure of the $r$-process of
heavy element nucleosynthesis in supernovae.  It could also aid the $p$-process
nucleosynthesis and supernova blow-up at an earlier stage in the supernova
process.  Finally the $\nu_s$ could bring about concordance in the present
discrepancy between the primordial abundance of $^4$He and the D/H ratio,
probably especially with nearly mass degenerate $\nu_\mu$ and $\nu_\tau$.  Both
the beneficial effects for the $r$-process and for big-bang nucleosynthesis
require a large $\Delta m^2_{e\mu}$, compatible with that needed for dark
matter.

Despite the appeal of being able to explain so many things with this mass
pattern, there is reluctance among some to accept light sterile neutrinos. 
Theoretically, sterile neutrinos are quite usual.  Indeed, the most natural is
to have one for each generation, but generally these are very heavy, and the
main problem is to make at least one of them light.  Several schemes have been
suggested, and a particularly appealing one is that by
Langacker\hphantom{,}\cite{ref:40} which occurs in a class of string models.

Finally, a warning: this otherwise desirable neutrino mass pattern should lead
to negative experimental results which could have a stultifying effect on this
field.  The KARMEN experiment\hphantom{,}\cite{ref:20} is touted as being able
to check the LSND result, and it can do so except in the crucial $\sim6$ eV$^2$
region.  Since its source-to-detector distance is half that of LSND, this
$\Delta m^2$ is at an oscillation minimum, as is clear in Fig.~\ref{fig:9}(b). 
The null result from KARMEN would soon be followed by an even more devastating
negative result from SNO.  The SNO comparison of charge-current events to
neutral-current events would be consistent with no neutrino oscillation, which
is also what should be expected from a solar $\nu_e\to\nu_s$ transition.  The
CHORUS and NOMAD experiments, designed for single-neutrino dark matter, are
sensitive in the wrong $\Delta m^2$ region for $\nu_\mu\to\nu_\tau$ and will
get a null result, although a search for $\nu_e\to\nu_\tau$ could be
interesting.  If presently planned long-baseline oscillation experiments
proceed, it is possible that they will also fail to show positive results,
since the atmospheric results from Super-Kamiokande may indicate a $\Delta m^2$
an order of magnitude smaller than that for which the experiments were
originally designed.  The subject of neutrino mass has had a bad history, and
these negative results could lead to total disbelief in earlier positive
results and likely withdrawal of support for future work.

The situation can be helped by anticipation of these null results---and that is
the main point of this paper---and by doing the difficult experiments which
could get the field out of trouble.  First, the energy spectrum of solar
neutrinos, as in Fig.~\ref{fig:1}(b), and the angular distribution of atmospheric
$\nu_\mu/\nu_e$ must be measured as well as possible to
provide convincing evidence of oscillations.  Second, the $\Delta
m^2_{e\mu}\sim6$--8 eV$^2$ region of LSND must be checked.  It would be highly
desirable to change the source-to-detector distance in that experiment, but an
independent measurement would be even better.  We are on the verge of a
significant leap forward in understanding a wide range of phenomena, making it
essential to avoid this possible blocking of progress.

\section*{Acknowledgments}

Appreciation is due D.~Bauer, G.~Fiorentini, N.~Hata, E.~Kearns, C.~Lane,
G.~McLaughlin, B.~Ricci, R.~Svoboda, and S.~Yellin for supplying figures.  It
is a pleasure to thank G.M.~Fuller, R.N.~Mohapatra, J.R.~Primack, and
Y.-Z.~Qian for extensive contributions to different parts of this paper.  The
U.S.~Department of Energy is thanked for partial support of this work.  Special
thanks goes to L.~Roszkowski for the invitation to give this talk at his
excellent COSMO-97 conference.

\end{document}